\documentclass[12pt]{article}
\usepackage{epsf}
\usepackage{amsmath, amssymb}

\begin{document}

\newcommand{\rem}[1]{{\bf #1}}

\renewcommand{\thefootnote}{\fnsymbol{footnote}}
\setcounter{footnote}{0}
\begin{titlepage}

\def\thefootnote{\fnsymbol{footnote}}

\begin{center}

\hfill UT-11-26\\
\hfill IPMU-11-0129\\
\hfill August, 2011\\

\vskip .75in

{\Large \bf 
Non-anomalous Discrete $R$-symmetry, Extra Matters,
and Enhancement of the Lightest SUSY Higgs Mass
}

\vskip .75in

{\large
Masaki Asano$^{(a)}$, Takeo Moroi$^{(a,b)}$, Ryosuke Sato$^{(a,b)}$
\\
and Tsutomu T. Yanagida$^{(a,b)}$
}

\vskip 0.25in

{\em $^{(a)}$Department of Physics, University of Tokyo,
Tokyo 113-0033, Japan
}

\vskip 0.1in

{\em $^{(b)}$Institute for the Physics and Mathematics of the Universe,\\
University of Tokyo, Kashiwa 277-8568, Japan}

\end{center}
\vskip .5in

\begin{abstract}

  We consider low-energy supersymmetric model with non-anomalous
  discrete $R$-symmetry.  In such a model, to make the $R$-symmetry
  non-anomalous, new particles with gauge quantum numbers should be
  inevitably added to the particle content of the minimal
  supersymmetric standard model (MSSM).  Those new particles may
  couple to the Higgs boson, resulting in a significant enhancement of
  the lightest Higgs mass.  We show that, in such a model, the
  lightest Higgs mass can be much larger than the MSSM upper bound;
  the lightest Higgs mass as large as $140\ {\rm GeV}$ (or larger)
  becomes possible.

\end{abstract}

\end{titlepage}

\renewcommand{\thepage}{\arabic{page}}
\setcounter{page}{1}
\renewcommand{\thefootnote}{\#\arabic{footnote}}
\setcounter{footnote}{0}

\section{Introduction}

If supersymmetry (SUSY) survives at low-energy scale ($\sim 1$ TeV),
discrete $R$-symmetries $Z_{NR}$ (with $N>2$) seem to play an
important role. First of all, the constant term in superpotential,
that is the gravitino mass, is only controlled by the discrete
$R$-symmetries. Therefore, the low-scale breaking of the discrete
$R$-symmetries may account for the presence of SUSY at TeV scale, for
example. The discrete $R$-symmetries may play a role of suppressing
dangerous dimension 4 operators for proton decays and they also may
guarantee the required long lifetime of the lightest SUSY particle as
a dark matter candidate.\footnote
{When the discrete $Z_{NR}$ (with $N>2$) is broken down to the
  $R$-parity $Z_{2R}$, we may have too many domain walls. For
  solutions to this domain wall problem, see \cite{Dine:2010eb}.}
It is known that approximate continuous
$R$-symmetries seem an essential component of dynamical SUSY
breaking. Those approximate $U(1)_R$ symmetries might be realized as
an accidental effective symmetry of the discrete $R$-symmetries
$Z_{NR}$ with $N>2$.

However, the discrete $Z_{NR}$ symmetries have gauge anomalies in the
minimal SUSY standard model (MSSM) \cite{Ibanez:1991hv}. Thus, if it
is the case, there is no reason to assume such discrete symmetries as
an (almost) exact symmetries. Therefore it seems more interesting to
cancel the unwanted anomalies by adding extra matters in the MSSM. In
\cite{Kurosawa:2001iq} it is shown that the discrete $Z_{4R}$ symmetry
is one of a few candidates. In fact, all gauge anomalies of $Z_{4R}$
can be canceled out by adding a pair of ${\bf 5} +{\bf \bar{5}}$
chiral multiplets or a pair of ${\bf 10} +{\bf \bar{10}}$. Their
masses are predicted to be at the same order of the Higgsino mass, say
$\sim 1$ TeV. Thus, they must give a significant contribution to
low-energy physics.  In particular, in the latter solution, up-type
Higgs $H_u$ can couple to the extra matters in the ${\bf 10}$ multiplet
as $W\sim UQH_u$ (where $U$ and $Q$ have the same gauge quantum
numbers as right-handed up-type quarks and left-handed quark doublets,
respectively).

In this paper, we show that the mass of the lightest Higgs boson can
be raised up to $\sim 140\ {\rm GeV}$ because of the extra Yukawa
coupling even when the SUSY-breaking scale is 1 TeV. This will be
tested at LHC soon.

\section{Non-Anomalous Discrete $R$-symmetry}

In this section, following \cite{Kurosawa:2001iq}, we discuss
anomaly-free conditions of discrete $R$-symmetry, $Z_{NR}$, in the
framework of $SU(5)$ Grand Unified Theories (GUTs).  We assume that
neutrino masses are explained by seesaw mechanism
\cite{Yanagida:1979as, GellMann:1976pg, Minkowski:1977sc}. Then, the
superpotential of the minimal SUSY $SU(5)$ GUT is of the following
form:
\begin{eqnarray}
  W_{\rm GUT} \sim \Phi_{\bf 10} \Phi_{\bf 10} H 
  + \Phi_{\bf 10} \bar{\Phi}_{\bf \bar{5}} \bar{H} 
  + \bar{\Phi}_{\bf \bar{5}} \bar{N}\bar{H}
  + \frac{1}{2} M_N \bar{N}\bar{N} 
  + \mu_H H\bar{H},
  \label{eq:W_MSSM}
\end{eqnarray}
where $M_N$ is the Majorana mass of right-handed neutrinos.  The
quantum numbers of the fields are shown in Table \ref{tab:SU5}.

\renewcommand{\arraystretch}{1.3}
\begin{table*}[t]
  \center{
    \begin{tabular}{|c||c|c|c|c|c|}
      \hline
      & $\Phi_{\bf 10}$ & $\bar{\Phi}_{\bf \bar{5}}$ 
      & $\bar{N}$ & $H$ & $\bar{H}$
      \\ \hline
      $SU(5)_{\rm GUT}$ 
      & {\bf 10} & ${\bf \bar{5}}$ & {\bf 1}
      & {\bf 5}  & ${\bf \bar{5}}$
      \\ \hline
      $Z_{NR}$  
      & $\phi_{\bf 10}$ & $\bar{\phi}_{\bf \bar{5}}$ & $\bar{\nu}$
      & $h$      & $\bar{h}$ \\
      \hline
    \end{tabular}
  }
  \caption{The matter content of the supersymmetric $SU(5)$ GUT, 
    and the quantum numbers of the fields.
    The $Z_{NR}$ charge of the Grassmann coordinate, $\theta$, 
    is denoted as $\alpha$.}
  \label{tab:SU5}
\end{table*}
\renewcommand{\arraystretch}{1}

To make the model invariant under the discrete $R$-symmetry, the
$Z_{NR}$ charges should satisfy the following conditions:
\begin{eqnarray}
  2 \phi_{\bf 10} + h &=& 2 \alpha \quad {\rm mod} ~N, \\ 
  \label{eq:W_1}
  \phi_{\bf 10} + \bar{\phi}_{\bf \bar{5}} + \bar{h} &=& 
  2 \alpha \quad {\rm mod} ~N, \\ 
  \label{eq:W_2}
  \bar{\phi}_{\bf \bar{5}} + \bar{\nu} + h &=& 
  2 \alpha \quad {\rm mod} ~N, \\ 
  \label{eq:W_3}
  2 \bar{\nu} &=& 2 \alpha \quad {\rm mod} ~N.
  \label{eq:W_4}
\end{eqnarray}
Here, we assume that the $\mu_H$-term is generated by the
Giudice-Masiero mechanism \cite{Giudice:1988yz}, and that
$Z_{NR}$-symmetry prevents $\mu_H$-parameter from being Planck
scale. Then, the following conditions are imposed:
\begin{eqnarray}
  h + \bar{h} = 0 \quad {\rm mod} ~N, ~~~\mbox{and}~~~
  h + \bar{h} \neq 2 \alpha \quad {\rm mod} ~N,
  \label{eq:mu}
\end{eqnarray}
which reduce to
\begin{eqnarray}
  2\alpha \neq 0 \quad {\rm mod} ~N.
  \label{2alpha_neq_0}
\end{eqnarray}

Next, we consider the conditions for anomaly cancellation
\cite{Ibanez:1991hv}.  For $Z_{NR}[SU(3)_C]^2$ and
$Z_{NR}[SU(2)_L]^2$:
\begin{eqnarray}
  Z_{NR}[SU(3)_C]^2 &:& \frac{3}{2} \left\{ 3 (\phi_{\bf 10} - \alpha) 
    + (\bar{\phi}_{\bf \bar{5}} - \alpha) \right\} 
  + 3 \alpha = \frac{N}{2}k, \\
  Z_{NR}[SU(2)_L]^2 &:& \frac{3}{2} \left\{ 3 (\phi_{\bf 10} - \alpha) 
    + (\bar{\phi}_{\bf \bar{5}} - \alpha)  \right\} 
  +\frac{1}{2} \left\{   (h - \alpha) 
    +  (\bar{h} - \alpha)  \right\} 
  + 2 \alpha = \frac{N}{2}k^{\prime},
\end{eqnarray}
where $k$ and $k^\prime$ are integers. Using Eqs.\ (\ref{eq:W_1}),
(\ref{eq:W_2}) and (\ref{eq:mu}), these condition are rewritten as
\begin{eqnarray}
  Z_{NR}[SU(3)_C]^2 &:& 3 \alpha = \frac{N}{2}k, 
  \label{eq:A_SU3} \\
  Z_{NR}[SU(2)_L]^2 &:&   \alpha = \frac{N}{2}k^{\prime}.
  \label{eq:A_SU2}
\end{eqnarray}
Importantly, Eq.\ (\ref{eq:A_SU2}) contradicts with the condition
(\ref{2alpha_neq_0}).  Thus, an additional contribution from extra
matters is needed to realize a non-anomalous discrete $R$-symmetry in
the framework of $SU(5)$ GUT.

Because the extra matters have gauge quantum numbers, they have to be
heavy enough to avoid direct search constraints.  If the extra matters
are vector-like, their masses can be generated by the Giudice-Masiero
mechanism.  If so, their masses are expected to be around the mass
scale of MSSM superparticles (i.e., $\sim 1\ {\rm TeV}$).  In order
for the Giudice-Masiero mechanism to work, the following condition
should be satisfied:\footnote
{If SUSY invariant masses for them are allowed, their $R$ charges
  should satisfy $\phi' +\bar{\phi}'=2\alpha$. In this case they do
  not contribute to the anomalies.}
\begin{eqnarray}
  \phi^\prime + \bar{\phi}^\prime = 0 \quad {\rm mod} ~N,
  \label{eq:GMforPhiPrime}
\end{eqnarray}
where $\phi^\prime$ and $\bar{\phi}^\prime$ are the $Z_{NR}$ charge of
the extra matter multiplets $\Phi^\prime$ and $\bar{\Phi}^\prime$,
respectively.

Extra matters should be embedded in complete multiplets of the GUT
group $SU(5)_{\rm GUT}$ in order not to spoil the gauge coupling
unification.  In addition, requiring the perturbativity of the gauge
coupling constants up to the GUT scale, we cannot introduce too many
extra matters.  If the masses of extra matters are at $\sim 100\ {\rm
  GeV} - 1\ {\rm TeV}$, only a limited number of ${\bf 5}+{\bf
  \bar{5}}$ and/or ${\bf 10}+{\bf \bar{10}}$ pairs can be introduced;
if a larger representation of $SU(5)_{\rm GUT}$ is added at
$\mu\lesssim 1\ {\rm TeV}$, the gauge couplings become
non-perturbative below the GUT scale.  Denoting the numbers of ${\bf
  5}+{\bf \bar{5}}$ and ${\bf 10}+{\bf \bar{10}}$ pairs as $n_{{\bf
    5}'}$ and $n_{{\bf 10}'}$, respectively, perturbativity of the
gauge couplings requires
\begin{eqnarray}
  n_{{\bf 5}'}+3n_{{\bf 10}'}\leq 4.
  \label{nextra_max}
\end{eqnarray}
In addition, using Eq.\ \eqref{eq:GMforPhiPrime}, the conditions for
the anomaly cancellation are
\begin{eqnarray}
  Z_{NR}[SU(3)_C]^2 &:& 
  (3-n_{{\bf 5}'}-3n_{{\bf 10}'}) \alpha = \frac{N}{2}k, 
  \label{eq:A_SU3_general} \\
  Z_{NR}[SU(2)_L]^2 &:& 
  (1-n_{{\bf 5}'}-3n_{{\bf 10}'}) \alpha = \frac{N}{2}k^{\prime}.
  \label{eq:A_SU2_general}
\end{eqnarray}

The conditions \eqref{2alpha_neq_0}, \eqref{nextra_max},
\eqref{eq:A_SU3_general} and \eqref{eq:A_SU2_general} are
simultaneously satisfied only when $(n_{{\bf 5}'},n_{{\bf
    10}'})=(1,0)$, $(3,0)$, or $(0,1)$.  In these cases, $N$ should be
$4$ or $20$ \cite{Kurosawa:2001iq}; for $N=4$ and $20$, there exist
consistent charge assignments.  For $N=4$, for example, one may take
$(\phi_{\bf 10}, \bar{\phi}_{\bf \bar{5}}, \bar{\nu}, h, \bar{h},
\alpha) = (1,1,1,0,0,1)$.  

Among three possibilities, we are interested in the case of $(n_{{\bf
    5}'},n_{{\bf 10}'})=(0,1)$ because the newly introduced ${\bf 10}$
multiplet may couple to the up-type Higgs boson if its $Z_{NR}$ charge
(denoted as $\phi_{\bf 10}^\prime$) is equal to that of $\Phi_{\bf
  10}$.  Such a charge assignment does not conflict with any of the
conditions because $\phi_{\bf 10}^\prime$ is arbitrary as far as Eq.\
\eqref{eq:GMforPhiPrime} is satisfied.  In the following, we
concentrate on the case with an extra pair of ${\bf 10}+{\bf
  \bar{10}}$ multiplet, and study the mass of the lightest Higgs boson
in such a case.

\section{Higgs Mass}

Now we discuss the lightest Higgs mass, paying particular attention to
the contributions of loop diagrams with extra matters inside the loop.
As discussed in the previous section, some of the fields contained in
$\Phi_{\bf 10}^\prime$ may couple to up-type Higgs $H_u$ if it has
proper $Z_{NR}$ charge.\footnote
{Because the extra matters couple to the Higgs boson, one should care
  about the oblique corrections (i.e., so-called $S$- and
  $T$-parameters) due to these fields.  In the present case, the
  dominant contribution to the masses of these new particles is from
  the gauge-invariant operators and hence the oblique corrections
  become suppressed as these extra particles become heavy.  The
  oblique corrections become small enough if the new particles are as
  heavy as $\sim 1\ {\rm TeV}$; for more detail, see
  \cite{EvaIbeYan}.}
Then, if its Yukawa interaction is large, we expect a sizable
correction to the lightest Higgs mass as in the case of the top and
stop \cite{Okada:1990vk, Haber:1990aw, Ellis:1990nz}.  

To study the Higgs mass with extra matters, we first decompose
$\Phi_{\bf 10}^\prime$ and $\bar{\Phi}_{\bf \bar{10}}^\prime$ as
$\Phi_{\bf 10}^\prime=Q+U+E$ and $\bar{\Phi}_{\bf
  \bar{10}}^\prime=\bar{Q}+\bar{U}+\bar{E}$, where $Q({\bf 3},{\bf
  2},1/6)$, $U({\bf \bar{3}},{\bf 1},-2/3)$, $E({\bf 1},{\bf 1},1)$,
$\bar{Q}({\bf \bar{3}},{\bf 2},-1/6)$, $\bar{U}({\bf 3},{\bf 1},2/3)$,
and $\bar{E}({\bf 1},{\bf 1},-1)$ are gauge eigenstates of the
standard-model gauge group.  (The gauge quantum numbers for $SU(3)_C$,
$SU(2)_L$ and $U(1)_Y$ are shown in the parenthesis.)  Then, the
relevant part of the superpotential is given by\footnote
{For simplicity, we neglect the effects of possible CP violating
  phases in the new interaction terms; parameters $y_U$, $M_Q$, $M_U$,
  and $A_U$ are all taken to be real.}
\begin{eqnarray}
  W = y_t t_R^c q_L H_u
  + y_U U Q H_u
  + M_U \bar{U} U
  + M_Q \bar{Q} Q,
  \label{W}
\end{eqnarray}
and the soft SUSY breaking terms are\footnote
{For simplicity, we assume that the bi-linear SUSY breaking terms for
  extra matters are negligible.}
\begin{eqnarray}
  {\cal L}_{\rm soft} &=& 
  m_{\tilde{q}}^2 |\tilde{q}_L|^2
  + m_{\tilde{t}}^2 |\tilde{t}_R^c|^2
  + m_{\tilde{Q}}^2 |\tilde{Q}|^2
  + m_{\tilde{\bar{Q}}}^2 |\tilde{\bar{Q}}|^2
  + m_{\tilde{U}}^2 |\tilde{U}|^2
  + m_{\tilde{\bar{U}}}^2 |\tilde{\bar{U}}|^2
  \nonumber \\ &&
  + (y_t A_t \tilde{t}_R^c \tilde{q}_L H_u
  + y_U A_U \tilde{U} \tilde{Q} H_u
  + {\rm h.c.}),
  \label{Lsoft}
\end{eqnarray}
where $q_L({\bf 3},{\bf 2},1/6)$ and $t_R^c({\bf \bar{3}},{\bf
  1},-2/3)$ are standard-model quarks in the third generation, which
contain left- and right-handed top (s)quarks, respectively.  In
addition, the ``tilde'' is for superparticles.  (So, $\tilde{Q}$ is
the scalar component in the superfield $Q$, for example.)

We presume that the $M_Q$- and $M_U$-parameters are generated by the
Giudice-Masiero mechanism, so the masses of extra matters are expected
to be as heavy as the MSSM superparticles.  Then assuming a little
hierarchy between the electroweak scale and the masses of
superparticles, which is suggested by the sparticle search
experiments, we estimate the Higgs mass by using the effective field
theory approach.  Then, the relevant theory describing the energy
scale above $M_{\rm SUSY}$ (which is taken to be the ``typical'' mass
of superparticles) is the MSSM with extra matters, while the
low-energy effective theory below $M_{\rm SUSY}$ is the standard
model. Two theories should be matched at $\mu=M_{\rm SUSY}$ (with
$\mu$ being the renormalization scale).

In our analysis, we consider the case that only one light Higgs
doublet, which we call the standard-model-like Higgs doublet $H_{\rm
  SM}$, remains below $M_{\rm SUSY}$, which is consistent with the
assumption that the low-energy effective theory below $M_{\rm SUSY}$
is the standard model.  The potential of $H_{\rm SM}$ is denoted as
\begin{eqnarray}
  V_{\rm SM} = m_H^2 |H_{\rm SM}|^2 + \frac{1}{2} \lambda |H_{\rm SM}|^4.
\end{eqnarray}
For the calculation of the lightest Higgs mass, we need to know the
coupling constant $\lambda$ at $M_{\rm SUSY}$; once $\lambda(M_{\rm
  SUSY})$ is known, Higgs mass is estimated as\footnote
{For the calculation of $\lambda (m_h)$, the most important effect
  below $M_{\rm SUSY}$ is from the Yukawa interaction with the top
  quark.  Thus, for our renormalization group analysis, we stop the
  running of $\lambda$ at $\mu=m_t$.}
\begin{eqnarray}
  m_h^2 = \lambda (m_h) v^2,
\end{eqnarray}
where $v\simeq 246\ {\rm GeV}$ is the vacuum expectation value of the
standard-model-like Higgs boson.  Notice that $\lambda (m_h)$ is
related to $\lambda(M_{\rm SUSY})$ by solving renormalization group
equation in the framework of the standard model.\footnote
{In our numerical analysis, we use the top-quark mass of $m_t^{\rm
    (pole)}=172.9\ {\rm GeV}$ \cite{Nakamura:2010zzi}.  The pole mass
  is related to the $\overline{\rm MS}$ mass as \cite{Arason:1991ic}
  \begin{eqnarray*}
    \frac{m_t^{\rm (pole)}}
    {m_t^{\rm (\overline{\rm MS})}(m_t^{\rm (pole)})} =
    1 + \frac{4}{3}\frac{\alpha_s (m_t^{\rm (pole)})}{\pi}.
  \end{eqnarray*}
}

In the present case, $\lambda(M_{\rm SUSY})$ is given by
\begin{eqnarray}
  \lambda(M_{\rm SUSY}) = \frac{1}{4} (g_2^2 + g_1^2) \cos^2 2\beta
  + \delta \lambda_{\tilde{t}} + \delta \lambda' ,
\end{eqnarray}
where $g_2$ and $g_1$ are gauge coupling constants for $SU(2)_L$ and
$U(1)_Y$, respectively.  In addition, $\delta\lambda_{\tilde{t}}$ is
the threshold correction at $M_{\rm SUSY}$ due to the stop loop
diagram, while $\delta\lambda'$ is that from diagrams with extra
matters inside the loop.  Taking $m_{\tilde{q}}=m_{\tilde{t}}$ for
simplicity, we obtain \cite{Okada:1990gg}
\begin{eqnarray}
  \delta \lambda_{\tilde{t}} = \frac{3 y_t^4 \sin^4\beta}{8\pi^2}
  \left( \frac{A_t^2}{m_{\tilde{t}}^2}
    - \frac{A_t^4}{12m_{\tilde{t}}^4}
  \right).
  \label{dlamda_t}
\end{eqnarray}

We study the effect of the extra matters using one-loop contribution
to the effective potential:
\begin{eqnarray}
  \Delta V = 
  \Delta V^{\rm (B)} + \Delta V^{\rm (F)},
\end{eqnarray}
where $\Delta V^{\rm (B)}$ and $\Delta V^{\rm (F)}$ are contributions
of bosonic and fermionic loops, respectively.  $\Delta V^{\rm (B)}$ is
given by
\begin{eqnarray}
  \Delta V^{\rm (B)} = 
  \frac{3}{32\pi^2} {\rm Tr}
  \left[ ({\cal M}_{\rm B}^2 + \Delta {\cal M}_{\rm B}^2)^2 
    \left\{ 
      \ln  \left( \frac{{\cal M}_B^2 
          + \Delta {\cal M}_{\rm B}^2}{\mu^2} \right)
      - \frac{3}{2} \right\}
  \right],
\end{eqnarray}
where
\begin{eqnarray}
  {\cal M}_{\rm B}^2 = {\rm diag}
  (M_Q^2 + m_{\tilde{Q}}^2, M_Q^2 + m_{\tilde{\bar{Q}}}^2, 
  M_U^2 + m_{\tilde{U}}^2, M_U^2 + m_{\tilde{\bar{U}}}^2),
\end{eqnarray}
and
\begin{eqnarray}
  \Delta {\cal M}_{\rm B}^2 = \left( \begin{array}{cccc}
      y_U^2 |H_u|^2 & 0 & y_U A_U H_u^* & y_U M_U H_u^* \\
      0 & 0 & y_U M_Q H_u^* & 0 \\
      y_U A_U H_u & y_U M_Q H_u & y_U^2 |H_u|^2 & 0 \\
      y_U M_U H_u & 0 & 0 & 0 \\
    \end{array} \right).
\end{eqnarray}
Furthermore,
\begin{eqnarray}
  \Delta V^{\rm (F)} = 
  - \left. \Delta V^{\rm (B)} \right|_{A_U=
    m_{\tilde{Q}}^2=m_{\tilde{\bar{Q}}}^2
    =m_{\tilde{U}}^2=m_{\tilde{\bar{U}}}^2=0}.
\end{eqnarray}
Calculating the coefficient of $|H_u|^4$ term in $\Delta V$ and
replacing $H_u\rightarrow H_{\rm SM}\sin\beta$, $\delta\lambda'$ is
obtained.  Because $Q$ and $U$ are in a same multiplet of $SU(5)_{\rm
  GUT}$, the relation $M_Q=M_U$ holds at the GUT scale.  This equality
is violated by the renormalization group effect below the GUT scale, but
the most important effect, i.e., the QCD effect, does not spoil this
relation.  Thus, we adopt the approximation $M_Q=M_U$ (at $\mu=M_{\rm
  SUSY}$).  In addition, for simplicity, we approximate
$m_{\tilde{Q}}^2=m_{\tilde{\bar{Q}}}^2=m_{\tilde{U}}^2=m_{\tilde{\bar{U}}}^2$.
Then, $\delta\lambda'$ is given by
\begin{eqnarray}
  \delta\lambda' &=&
  \frac{3 y_U^4\sin^4\beta}{8\pi^2}
  \ln \left( \frac{M_U^2 + m_{\tilde{U}}^2}{M_U^2} \right)
  \nonumber \\ &&
  - \frac{y_U^4\sin^4\beta}{32\pi^2}
  \frac{A_U^4 - (8 M_U^2 + 12 m_{\tilde{U}}^2)A_U^2
    + 8 M_U^2 m_{\tilde{U}}^2 + 10 m_{\tilde{U}}^4}
  {(M_U^2 + m_{\tilde{U}}^2)^2}.
  \label{dlambda'}
\end{eqnarray}
One can see that, with a relevant choice of parameters,
$\delta\lambda'$ becomes positive and an enhancement of the lightest
Higgs mass happens.

To see how large the Higgs mass can be, we calculate $m_h$.  To make
our discussion simple, we take
\begin{eqnarray}
  m_{\tilde{t}}^2 = m_{\tilde{q}}^2
  = m_{\tilde{Q}}^2 = m_{\tilde{\bar{Q}}}^2
  = m_{\tilde{U}}^2 = m_{\tilde{\bar{U}}}^2
  \equiv m_{\rm SUSY}^2,
\end{eqnarray}
and the tri-linear coupling constants are parametrized as
\begin{eqnarray}
  A_t = a_t m_{\rm SUSY},~~~
  A_U = a_U m_{\rm SUSY}.
\end{eqnarray}
For our numerical calculation, we take $M_{\rm SUSY}=m_{\rm SUSY}$.

\begin{figure}
  \centerline{\epsfxsize=0.8\textwidth\epsfbox{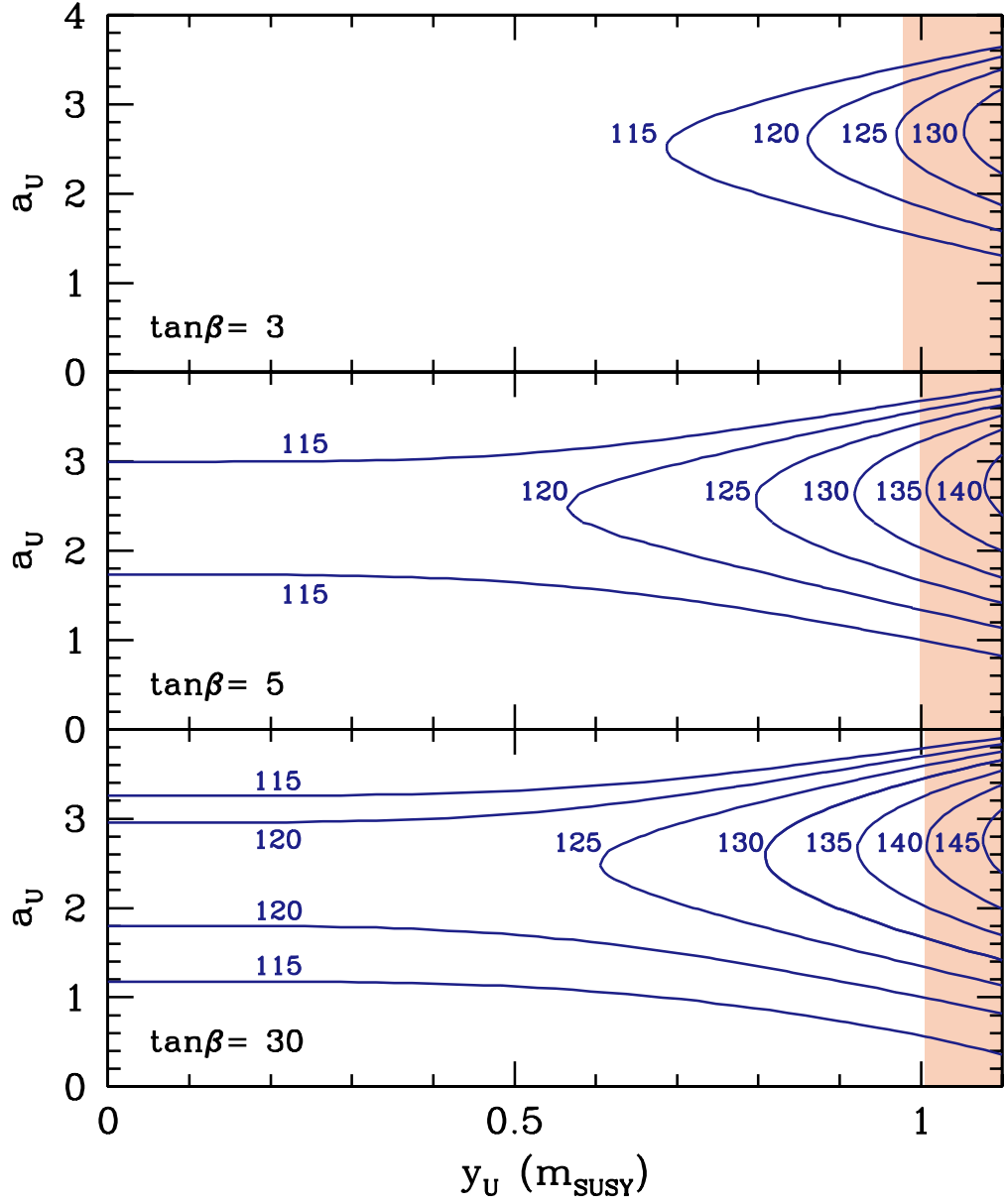}}
  \caption{\small Contours of constant $m_h$ on $y_U$ vs.\ $a_U$ plane
    for $\tan\beta=3$, $5$, and $30$.  Here, we have taken $m_{\rm
      SUSY}=M_U=M_Q=1\ {\rm TeV}$, and $a_t=a_U$.  In the shaded
    region, $y_U$ becomes non-perturbative below the GUT scale.
    Numbers in the figure are the lightest Higgs mass in units of GeV.}
\label{fig:mh_atuniv}
\end{figure}

\begin{figure}
  \centerline{\epsfxsize=0.8\textwidth\epsfbox{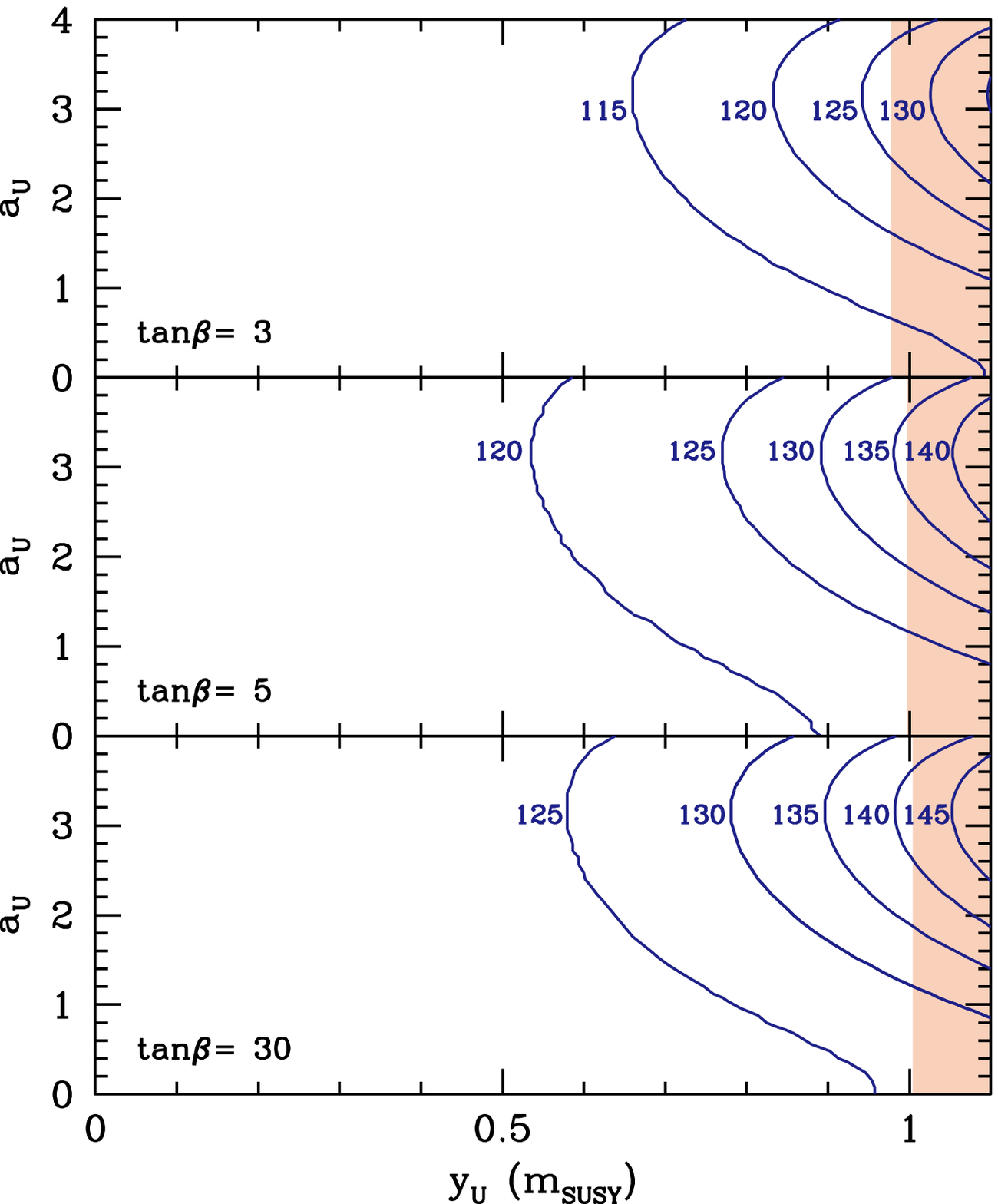}}
  \caption{\small Same as Fig.\ \ref{fig:mh_atuniv}, except for
    $a_t=\sqrt{6}$.}
\label{fig:mh_atmax}
\end{figure}

In Fig.\ \ref{fig:mh_atuniv}, we plot the lightest Higgs mass on $y_U$
(at $\mu=M_{\rm SUSY}$) vs.\ $a_U$ plane for several values of
$\tan\beta$, taking $m_{\rm SUSY}=1\ {\rm TeV}$, $M_U=M_Q=1\ {\rm
  TeV}$, and $a_t=a_U$.  If the Yukawa coupling constants are too
large, they diverge below the GUT scale.  Requiring the perturbativity
(i.e., $y_U^2\lesssim 4\pi$) below the GUT scale, the upper bound on
$y_U(m_{\rm SUSY})$ is obtained; in the figure, such a bound is also
shown.  In addition, as one can see from Eq.\ \eqref{dlamda_t},
$\delta\lambda_{\tilde{t}}$ takes its maximal value when
$a_t=\sqrt{6}$; results for such a case are shown in Fig.\
\ref{fig:mh_atmax}.

One can see that the radiative correction due to the extra particles
may drastically change the lightest Higgs mass; $m_h$ can be
significantly enhanced compared to the case of the MSSM
\cite{Moroi:1991mg, Moroi:1992zk}.  We note here that the lightest
Higgs mass is sensitive to the $A_U$-parameter.  In particular, when
$a_U\sim 3$ and $y_U\sim 1$, the lightest Higgs mass becomes as heavy
as $\sim 140\ {\rm GeV}$ even if we assume the perturbativity of the
Yukawa coupling constant up to the GUT scale.  Such a value of $m_h$
is above the MSSM bound on the lightest Higgs mass, which is $\sim
125\ {\rm GeV}$ if $m_{\rm SUSY}=1\ {\rm TeV}$ \cite{Carena:2002es}.
We have also checked that a larger value of $m_h$ is also possible for
a different value of $M_U$ or $m_{\tilde{U}}$.  As the ratio
$m_{\tilde{U}}^2/M_U^2$ becomes large, the logarithmic term in Eq.\
\eqref{dlambda'} is enhanced, resulting in a larger value of $m_h$.
For example, when $M_U=500-750\ {\rm GeV}$ (and $m_{\rm SUSY}=1\ {\rm
  TeV}$), $m_h$ can be made as large as $145-150\ {\rm GeV}$.

\section{Discussion}

In this paper, we have discussed the lightest Higgs mass in a model
with a non-anomalous discrete $R$-symmetry.  For the cancellation of
the gauge anomaly, extra particles should be added to the MSSM; we
have seen that the gauge anomaly can be cancelled out by adding ${\bf
  10}+{\bf \bar{10}}$ multiplet of $SU(5)_{\rm GUT}$.  In such a
model, the SUSY invariant mass term arises for the ${\bf 10}+{\bf
  \bar{10}}$ multiplet via the Giudice-Masiero mechanism, and the
particles in the ${\bf 10}+{\bf \bar{10}}$ multiplet becomes as light
as MSSM superparticles.  We have paid particular attention to the
lightest Higgs mass in such a model, and we have seen that $m_h$ can
become as large as $\sim 140\ {\rm GeV}$ (or larger).  

This fact has a great impact on the study of SUSY models because the
Higgs mass is the crucial check point of low-energy SUSY and also
because the LHC experiment is expected to find Higgs boson in near
future.  We have shown that the significant enhancement of
the Higgs mass is possible if extra particles from ${\bf 10}$ multiplet
of $SU(5)_{\rm GUT}$ exist; such a modification is well-motivated to
realize a non-anomalous discrete $R$-symmetry.  In particular, the
ATLAS group recently observed $\sim 2.8\sigma$ excess of the
Higgs-like events in the mass range of $\sim 120-140\ {\rm GeV}$
\cite{ATLAS-CONF-2011-112}.  If the existence of the Higgs boson
heavier than the MSSM bound is confirmed, it is strongly suggested to
look for extra particles in ${\bf 10}+{\bf \bar{10}}$ multiplet.

Our estimation of the Higgs mass is based on the renormalization-group
analysis with taking account of the leading-order threshold correction
at $\mu=M_{\rm SUSY}$; the sub-dominant contributions are expected to
be suppressed by powers of $v/M_{\rm SUSY}$ or $v/M_{U}$.  Such
sub-dominant contributions may slightly change the lightest Higgs
mass.  For more precise determination of $m_h$, the full one-loop
calculation of the effective potential is needed, which is beyond the
scope of this paper.  However, in order to estimate the accuracy of
our results, we have compared our results (for the case without
extra matter) with those of FeynHiggs package \cite{Heinemeyer:1998yj,
  Heinemeyer:1998np, Degrassi:2002fi, Frank:2006yh} which is expected
perform a precise calculation of the Higgs mass in the framework of
the MSSM.  We found that the difference between two results are
within $\sim 5\ {\rm GeV}$.

Before closing this paper, several comments are in order.  First
comment is on the stability of extra particles.  If we strictly adopt
the superpotential given in Eq.\ \eqref{W} and soft SUSY breaking
terms given in Eq.\ \eqref{Lsoft}, the lightest extra particle becomes
stable.  If a charged or colored particle becomes stable, it may
conflict with cosmological constraints.  However, the extra particles
can decay into standard-model quarks or leptons (and weak boson) if
they slightly mix with standard-model particles.  Because the $Z_{NR}$
charges of the extra particles are same as or opposite to that of the
standard-model fermions, such mixing naturally exists.

We have seen that the enhancement of the lightest Higgs mass becomes
significant when $\tan\beta$ is large.  This fact has an advantage if
we take the muon $(g-2)$ anomaly seriously.  If we compare the face
values of the measured value of the muon anomalous magnetic moment
with the theoretical prediction, they have $3.3\sigma$ discrepancy
\cite{Hagiwara:2011af}.  In low-energy supersymmetric models, SUSY
contribution to the muon anomalous magnetic moment becomes sizable in
particular when $\tan\beta$ is large \cite{Chattopadhyay:1995ae,
  Moroi:1995yh}, which may be the origin of the muon $(g-2)$ anomaly.
In the present set up, thus the muon $(g-2)$ anomaly may be solved
with realizing the Higgs mass much larger than the MSSM upper bound.
This is a big contrast to the case with a singlet Higgs, i.e., the
so-called the next to the MSSM (NMSSM), which may also enhance the
lightest Higgs mass; in the NMSSM, $\tan\beta$ is required to be
relatively small for the enhancement of $m_h$ \cite{Ellwanger:2009dp}.

{\it Acknowledgments:} The authors thank J. Evans and M. Ibe for
useful discussion.  This work is supported by Grant-in-Aid for
Scientific research from the Ministry of Education, Science, Sports,
and Culture (MEXT), Japan, No.\ 22244021 (M.A., T.M., and T.T.Y.),
No.\ 22540263 (T.M.), and also by the World Premier International
Research Center Initiative (WPI Initiative), MEXT, Japan. The work of
R.S. is supported in part by JSPS Research Fellowships for Young
Scientists.


\begin{thebibliography}{99}

\bibitem{Dine:2010eb}
  M.~Dine, F.~Takahashi and T.~T.~Yanagida,
  JHEP {\bf 1007} (2010) 003.

\bibitem{Ibanez:1991hv}
  L.~E.~Ibanez and G.~G.~Ross,
  Phys.\ Lett.\  B {\bf 260} (1991) 291.

\bibitem{Kurosawa:2001iq}
  K.~Kurosawa, N.~Maru and T.~Yanagida,
  Phys.\ Lett.\  B {\bf 512} (2001) 203.

\bibitem{Yanagida:1979as}
  T. Yanagida,
  in {\it Proc, of the Workshop on the Unified Theory
    and Baryon Number in the Universe,} 
  ed. O. Sawada and A. Sugamoto,
  (KEK report 79-18, 1979).

\bibitem{GellMann:1976pg}
  M. Gell-Mann, P. Ramond and R. Slansky, 
  in {\it Supergravity},
  ed P. van Nieuwenhuizen and D.Z. Freedman,
  (North Holland, Amsterdam, 1979)

\bibitem{Minkowski:1977sc}
  P.~Minkowski,
  Phys.\ Lett.\  B {\bf 67} (1977) 421.

\bibitem{Giudice:1988yz}
  G.~F.~Giudice and A.~Masiero,
  Phys.\ Lett.\  B {\bf 206} (1988) 480.

\bibitem{EvaIbeYan}
  J. Evans, M. Ibe and T.T. Yanagida, in preparation.

\bibitem{Okada:1990vk}
  Y.~Okada, M.~Yamaguchi and T.~Yanagida,
  Prog.\ Theor.\ Phys.\  {\bf 85} (1991) 1.

\bibitem{Haber:1990aw}
  H.~E.~Haber and R.~Hempfling,
  Phys.\ Rev.\ Lett.\  {\bf 66} (1991) 1815.

\bibitem{Ellis:1990nz}
  J.~R.~Ellis, G.~Ridolfi and F.~Zwirner,
  Phys.\ Lett.\  B {\bf 257} (1991) 83.

\bibitem{Nakamura:2010zzi}
  K.~Nakamura {\it et al.}  [Particle Data Group],
  J.\ Phys.\ G {\bf 37} (2010) 075021.

\bibitem{Arason:1991ic}
  H.~Arason, D.~J.~Castano, B.~Keszthelyi, S.~Mikaelian, E.~J.~Piard, 
  P.~Ramond and B.~D.~Wright,
  Phys.\ Rev.\  D {\bf 46} (1992) 3945.

\bibitem{Okada:1990gg}
  Y.~Okada, M.~Yamaguchi and T.~Yanagida,
  Phys.\ Lett.\  B {\bf 262} (1991) 54.

\bibitem{Moroi:1991mg}
  T.~Moroi and Y.~Okada,
  Mod.\ Phys.\ Lett.\  A {\bf 7} (1992) 187.

\bibitem{Moroi:1992zk}
  T.~Moroi and Y.~Okada,
  Phys.\ Lett.\  B {\bf 295} (1992) 73.

\bibitem{Carena:2002es}
  M.~S.~Carena and H.~E.~Haber,
  Prog.\ Part.\ Nucl.\ Phys.\  {\bf 50} (2003) 63.

\bibitem{ATLAS-CONF-2011-112}
  The ATLAS collaboration, 
  ATLAS-CONF-2011-112 (July, 2011).

\bibitem{Heinemeyer:1998yj}
 S.~Heinemeyer, W.~Hollik and G.~Weiglein,
 Comput.\ Phys.\ Commun.\  {\bf 124} (2000) 76.

\bibitem{Heinemeyer:1998np}
 S.~Heinemeyer, W.~Hollik and G.~Weiglein,
 Eur.\ Phys.\ J.\  C {\bf 9} (1999) 343.

\bibitem{Degrassi:2002fi}
 G.~Degrassi, S.~Heinemeyer, W.~Hollik, P.~Slavich and G.~Weiglein,
 Eur.\ Phys.\ J.\  C {\bf 28} (2003) 133.

\bibitem{Frank:2006yh}
 M.~Frank, T.~Hahn, S.~Heinemeyer, W.~Hollik, H.~Rzehak and G.~Weiglein,
 JHEP {\bf 0702} (2007) 047.

\bibitem{Hagiwara:2011af}
  K.~Hagiwara, R.~Liao, A.~D.~Martin, D.~Nomura and T.~Teubner,
  J.\ Phys.\ G {\bf 38} (2011) 085003.

\bibitem{Chattopadhyay:1995ae}
  U.~Chattopadhyay and P.~Nath,
  Phys.\ Rev.\  D {\bf 53} (1996) 1648.

\bibitem{Moroi:1995yh}
  T.~Moroi,
  Phys.\ Rev.\  D {\bf 53} (1996) 6565
  [Erratum-ibid.\  D {\bf 56} (1997) 4424].

\bibitem{Ellwanger:2009dp}
  See, for example, U.~Ellwanger, C.~Hugonie and A.~M.~Teixeira,
  Phys.\ Rept.\  {\bf 496} (2010) 1.

\end{thebibliography}
\end{document}